\documentclass[a4paper]{article}

\usepackage[numbers]{natbib}

\usepackage[bf]{caption2}

\usepackage{amsmath}
\usepackage{graphicx}

\usepackage{txfonts}
\usepackage{hyperref}

\newcommand{\um}[1]{\ensuremath{\,\mathrm{#1}}}   % units of measure
\newcommand{\di}{\ensuremath{\,\mathrm{d}}}       % differential
\newcommand{\mtrx}[1]{\ensuremath{\mathsf{#1}}}   % matrix
\newcommand{\uvect}[1]{\ensuremath{\hat{\boldsymbol{#1}}}} % unit vector
\newcommand{\ps}{\ensuremath{\boldsymbol{\cdot}}}  % scalar product
\newcommand{\pv}{\ensuremath{\boldsymbol{\times}}} % vector product
\title{\textsf{\textbf{Sky coverage of orbital detectors.  Analytical
approach}}}

\author{\textbf{Diego Casadei}\\
\normalsize\emph{INFN, Sezione di Bologna, Via Irnerio 46, 40126
  Bologna, Italy}\\ 
\normalsize\url{casadei.diego@gmail.com}}

\date{\normalsize%
 Version 1.2 ---
December 28, 2005}

\begin{document}

\maketitle

\begin{abstract}
Orbital detectors without pointing capability have to keep their field
of view axis laying on their orbital plane, to observe the largest sky
fraction.  A general approach to estimate the exposure of each sky
element for such detectors is a valuable tool in the R\&D phase of a
project, when the detector characteristics are still to be fixed.  An
analytical method to estimate the sky exposure is developed, which
makes only few very reasonable approximations.  The formulae obtained
with this method are used to compute the histogram of the sky exposure
of a hypothetical gamma-ray detector installed on the ISS.  The C++
code used in this example is freely available on the
%\url{http://cern.ch/casadei/software.html} 
{\small\url{https://github.com/dcasadei/SkyCoverage}}
web page.
\end{abstract}

\section{Introduction}

 In addition to the large, complex and very expensive orbiting
 observatories, smaller size and lower cost detectors, whose goals are
 restricted to a narrower field, can perform interesting astrophysical
 measurements.  To be (relatively) cheap, a satellite cannot have
 special pointing capabilities, beyond the normal gyroscopic
 stabilization and control, which is necessary not to loose
 communication with Earth stations and to keep safe satellite
 orientation.

 When trying to define what are the parameters which should
 characterize an orbiting detector, one would like to be able to get
 values which are good estimations of the final parameters much before
 developing a complete Monte Carlo simulation.  As an example, let us
 consider a detector which aims to measure the cosmic gamma-ray
 background in a wide energy range: its design acceptance and the
 mission duration will depend on the expected flux of known sources,
 in addition to the detector energy resolution.

 To estimate the detector field of view and sensitive area, together
 with the mission duration, one needs to know the measured flux (per
 energy bin) coming from each sky element, which depends on the
 sources and on the detector characteristics.  Thus, one will first
 produce the histograms of the sky exposure for different mission
 durations and then will consider different sensitive areas (i.e.\
 different detector configurations), in order to find the
 configuration which best matches the expected fluxes.

 Another case in which a method to determine the sky exposure for a
 given detecor can be very useful is when one tries to understand the
 possible performance of an existing detector for the search for a new
 ``channel'' in the data analysis, and the access to the detailed
 simulation software is not easy, fast, or allowed.  If the obtained
 values suggest the possibility of doing interesting astrophysics, one
 will of course use the official detector simulation program to refine
 the estimation.

 In this paper, an analytical method to obtain the sky coverage of an
 orbiting detector is shown, which makes use of few reasonable
 assumptions to simplify the computation: (1) the orbit eccentricity
 is zero; (2) the orbit precession period is much larger than the
 revolution period; (3) the detector field of view is a cone, whose
 axis lies in the orbital plane, rotating with the same angular
 velocity of the satellite revolution.  These assumptions are valid in
 the case of a detector installed on the International Space
 Station\footnote{\url{http://www.hq.nasa.gov/osf/station/viewing/issvis.html}}
 (ISS), taken as an example in the following sections.  A program has
 been developed which makes use of this method: its C++ code is freely
 available on the \url{http://cern.ch/casadei/software.html} web page.

\section{The method assumptions and algorithm}\label{sec-hyp-alg}

\subsection{Hypotheses}

 The first hypothesis (null eccentricity $e$) implies that the
 revolution velocity is constant.  Its explicit dependence on the
 orbit parameters is shown in ref.\ \citep{murad95}: the result is a
 function with a rather weak dependence on $e$, hence this
 approximation is quite good for small values of $e$.

 The second hypothesis (orbit precession period $T \gg \tau$, where
 $\tau$ is the revolution period) is valid in general, and allows for
 a useful approximation: the method first computes the sky exposure
 over a single orbit, then rotates this map following the orbit
 precession.

 The hypothesis that the field of view is a cone with axis in the
 orbital plane is not true in general.  The real detector acceptance
 is usally different from a uniform cone, though one often accepts
 tracks within some ``fiducial'' cone to make easier the off-line data
 analysis (avoiding ``border effects'').  Having the axis contained in
 the orbital plane maximizes the fraction of the sky seen by the
 detector, hence it is a quite natural choice.  Finally, having
 angular velocity equal to that of the revolution motion means that
 the angle (in the orbital plane) between the cone axis and the radial
 vector (from the orbit center to the satellite) is kept constant.
 This is true for detectors installed on the ISS.

 The detector acceptance will in general depend both on the incident
 angle $\alpha$ of the photon and on its energy, with $\alpha=0$ for
 tracks parallel to the cone axis.  However, we neglect this angular
 dependency and consider a uniform acceptance in $\alpha$ inside a
 cone with half-aperture $\alpha_\mathrm{c}$.  The analytical method
 presented here is really fast\footnote{To compute the final histogram
 of this paper, shown in figure~\ref{grs-prec}, a Linux based PC with
 a Pentium III 1 GHz CPU takes 1.8 s.} from the computing point of
 view, and it is easy to change the source code to implement the
 effect of the $\alpha$ dependence of the acceptance (only a single
 file needs to be changed).  On the other hand, it will be much more
 tricky to implement the azimuthal dependence of the detector
 acceptance: if the user really needs to reach this level of detail,
 she/he would benefit from switching to the official detector
 simulation.

\subsection{Reference systems}

 In this paper, three reference systems are used: the orbit reference
 system (hereafter ORS), the galactic reference system (GRS), and the
 precession reference system (PRS).  The GRS is the system in which we
 usually want to express the results.  The PRS is used as intermediate
 step by the algorithm, to simplify the precession computation.
 Finally, the equatorial $(x,y)$ plane of the ORS is the plane of the
 orbit.  PRS and GRS are inertial reference sytems (they do not
 rotate).

 Because we will consider here only mission durations which are at
 least few times longer than the precession period\footnote{Increasing
 somewhat the mission duration is simpler and cheaper than increasing
 the detector sensitive area (i.e.\ the payload weight).}, the exact
 position of the detector when it starts data taking is not important.
 Hence, we are only interested in the orbit unit vector \uvect{n},
 which defines the $z$ axis in the ORS.  The $(x,y)$ plane is the
 orbital plane in this reference system, and we are free to choose the
 axes direction in ordert to simplify the computation.

 A possible choice for the reference axes $(X, Y, Z)$ in the GRS is
 the following.  The axes are defined by the three unit vectors
 \uvect{I}, \uvect{J} and \uvect{K} respectively: \uvect{I} points to
 the vernal equinox direction, \uvect{K} to the North Pole and
 \uvect{J} completes the right-handed triplet \citep{murad95}.  Any
 point in the sky can be addressed in the GRS using two angles:
 $\Theta \in [0, \pi]$, which is the angle formed with \uvect{K}, and
 $\Phi \in [0, 2\pi]$, expressing the rotation in the equatorial $(X,
 Y)$ plane, starting from the \uvect{I} unit vector.  Astronomers
 usually adopt the ``equatorial coordinates'' declination $\delta =
 \pi/2-\Theta$ and right ascension $\alpha = \Phi$ instead of $\Theta$
 and $\Phi$.  However, using the colatitude $\Theta$ instead of
 $\delta$ simplifies the histogramming.

 The orbit precession is described by the constant unit vector
 \uvect{k} about which the orbit normal \uvect{n} rotates while
 keeping $\uvect{k}\ps\uvect{n}(t) = \cos\beta$ constant.  The
 \uvect{k} unit vector defines the $z'$ axis of the PRS, and we are
 free to choose the direction of the $x'$ and $y'$ axes to simplify
 the computation.  The algorithm needs the PRS when computing the
 orbit precession effect on the sky exposure map, which in the PRS is
 simply a rotation around \uvect{k} of the histogram obtained for a
 single orbit.

\subsection{Algorithm}

 For each sky element, its exposure is equal to the product of its
 solid angle and the total time for which it is found inside the
 detector acceptance.  During few orbits, as long as
 $\di\uvect{n}/\di{t}$ can be neglected, the same points are seen by
 the detector few times: if $\alpha_\mathrm{c} < 90$ deg (as usual),
 the celestial sphere is divided in two parts by the ``belt'' viewed
 by the detector.  The exposure is largest for points situated along
 the central part of the belt, and decreases to zero going from the
 center to the boundaries.

 After a large number of orbits, the precession of the belt increases
 the sky fraction viewed by the detector.  For each sky element, its
 exposure is equal to the convolution between the time spent inside
 the belt (which rotates following the orbit precession) and the
 exposure function inside the belt (computed along one revolution).
 
 The algorithm, for the case in which the detector acceptance is
 uniform inside a cone with half-aperture $\alpha_\mathrm{c}$ whose
 axis is defined as the detector axis, is based on the following
 ideas:
\begin{enumerate}
 \item to find the exposure map of the sky in the ORS for one
       revolution, we will follow an analytical treatment
       (section~\ref{sec-ors});

 \item the (slow) orbit precession in the PRS is computed with a
       discrete algorithm acting on the exposure histogram
       (section~\ref{sec-prs});

 \item the exposure map of the sky in the GRS is found analytically
       (section~\ref{sec-grs}).
\end{enumerate}

 If the real detector acceptance does not have cylindrical symmetry,
 one has first to compute the real belt exposure along one orbit by
 changing step 1 (i.e.\ by changing a single C function in the code),
 and then can proceed as before.  When considering the dependence of
 the acceptance on the photon energy, one will repeat the three steps
 independently in each energy bin.

\section{Single orbit exposure map}\label{sec-ors}

\begin{figure}[t!]
 \centering\noindent
 \includegraphics[scale=0.45]{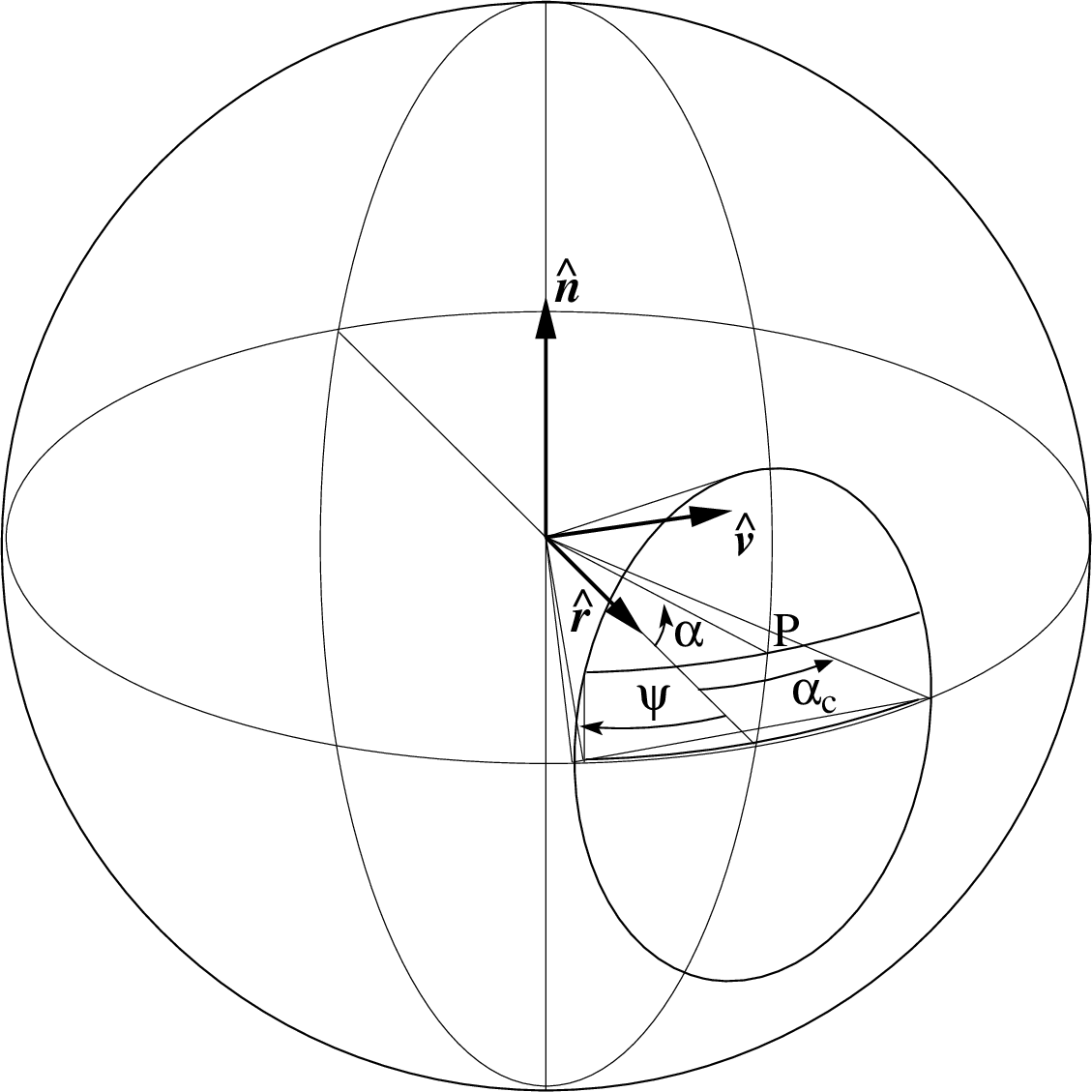}
 \caption{Sky sphere viewed by a cone.}\label{sphere}
\end{figure}

 Let the detector field of view be a cone of half-aperture
 $\alpha_\mathrm{c}$ (figure~\ref{sphere}), with axis along the unit
 vector \uvect{r} (the ``local vertical'')\footnote{If the detector
 axis, which lies in the $(x, y)$ plane, forms a constant angle with
 \uvect{r}, the result is exactly the same.}.  The latter turns around
 \uvect{n} with angular velocity $\omega = 2\pi/\tau$, and instant
 velocity unit vector $\uvect{v} = \uvect{n} \pv \uvect{r}$ (the
 ``local horizontal'')\footnote{The three unit vectors \uvect{n},
 \uvect{v} and \uvect{r}, form the basis of the so-called
 ``local-vertical local-horizontal'' (LVLH) reference system.}.

\subsection{Sky coverage in the orbit reference system}

 All points along the \uvect{r} trajectory are inside the cone for a
 time $t = 2 \alpha_\mathrm{c} / \omega = \alpha_\mathrm{c} \tau /
 \pi$ during each orbit.  The fraction of time during which a point P
 in the sky is found inside the cone depends on the angle $\alpha$
 formed with the equatorial plane (which contains \uvect{r} and
 \uvect{v}).  This time $t(\alpha)$ is maximum when $\alpha = 0$ and
 goes to zero when $\alpha$ approaches $\alpha_\mathrm{c}$.  For
 higher angles, the point is never contained by the cone (we restrict
 our analysis to $\alpha_\mathrm{c}<90$ deg).

\begin{figure}[t!]
\centering
\begin{minipage}{0.45\textwidth}
 \includegraphics[scale=0.45]{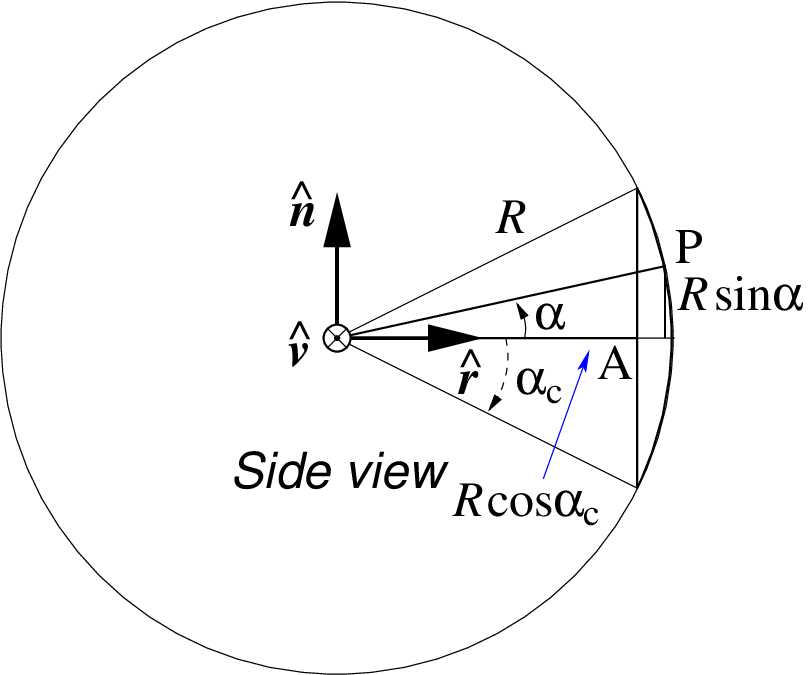}
 \caption{Side view of the sky sphere.}\label{spheres-side}
\end{minipage}
\hfill
\begin{minipage}{0.45\textwidth}
\centering
 \includegraphics[scale=0.45]{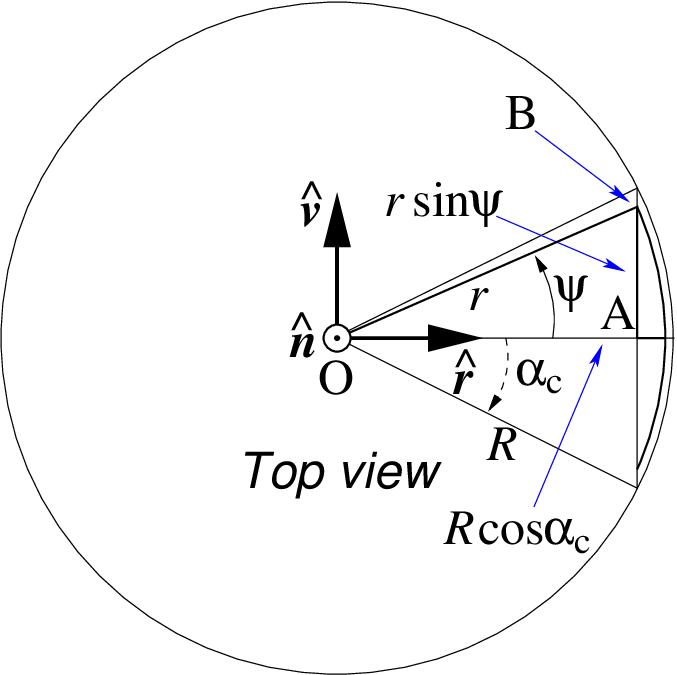}
 \caption{Top view of the sky sphere.}\label{spheres-top}
\end{minipage}%
\end{figure}

 The ``parallel'' passing through P is a circumference with length
 $\ell = 2\pi R \cos\alpha$ (figure~\ref{spheres-side}).  Let $\psi$
 be the half angular width of this parallel, when projected to the
 equatorial plane (figure~\ref{spheres-top}): when $\alpha$ approaches
 $\alpha_\mathrm{c}$, $\psi$ goes to zero; as $\alpha$ goes to zero,
 $\psi$ approaches $\alpha_\mathrm{c}$.  The exposure time $t(\alpha)$
 is equal to the ratio between the parallel width $2\psi$ and the
 angular velocity:
\begin{equation}
  t(\alpha) = \frac{2 \, \psi(\alpha)}{\omega}
            = \frac{\tau}{\pi} \, \psi (\alpha) \; .
\end{equation}

 In order to find $\psi(\alpha)$, we consider the triangle ABO on the
 equatorial plane (see figure~\ref{spheres-top}) and observe that the
 segment BO, the radius of the parallel passing through P, is $r = R
 \cos\alpha$.  The lengths of the segments AO and AB are
 $R\cos\alpha_\mathrm{c}$ and $r\sin\psi$ respectively.  The
 Pitagora's theorem implies that
\begin{equation}
  \cos^2\alpha = \cos^2\alpha_\mathrm{c} + 
                 \cos^2\alpha\sin^2\psi \; ,
\end{equation}
 hence one finds that $\sin\psi$ can be written as:
\begin{equation}
  \sin\psi = \sqrt{1 - \frac{\cos^2\alpha_\mathrm{c}}{\cos^2\alpha}}
  \; .
\end{equation}
 One can easily verify that this expression has the expected behavior
 when $\alpha$ approaches zero or $\alpha_\mathrm{c}$.

\begin{figure}[t]
\centering
\includegraphics[scale=0.35]{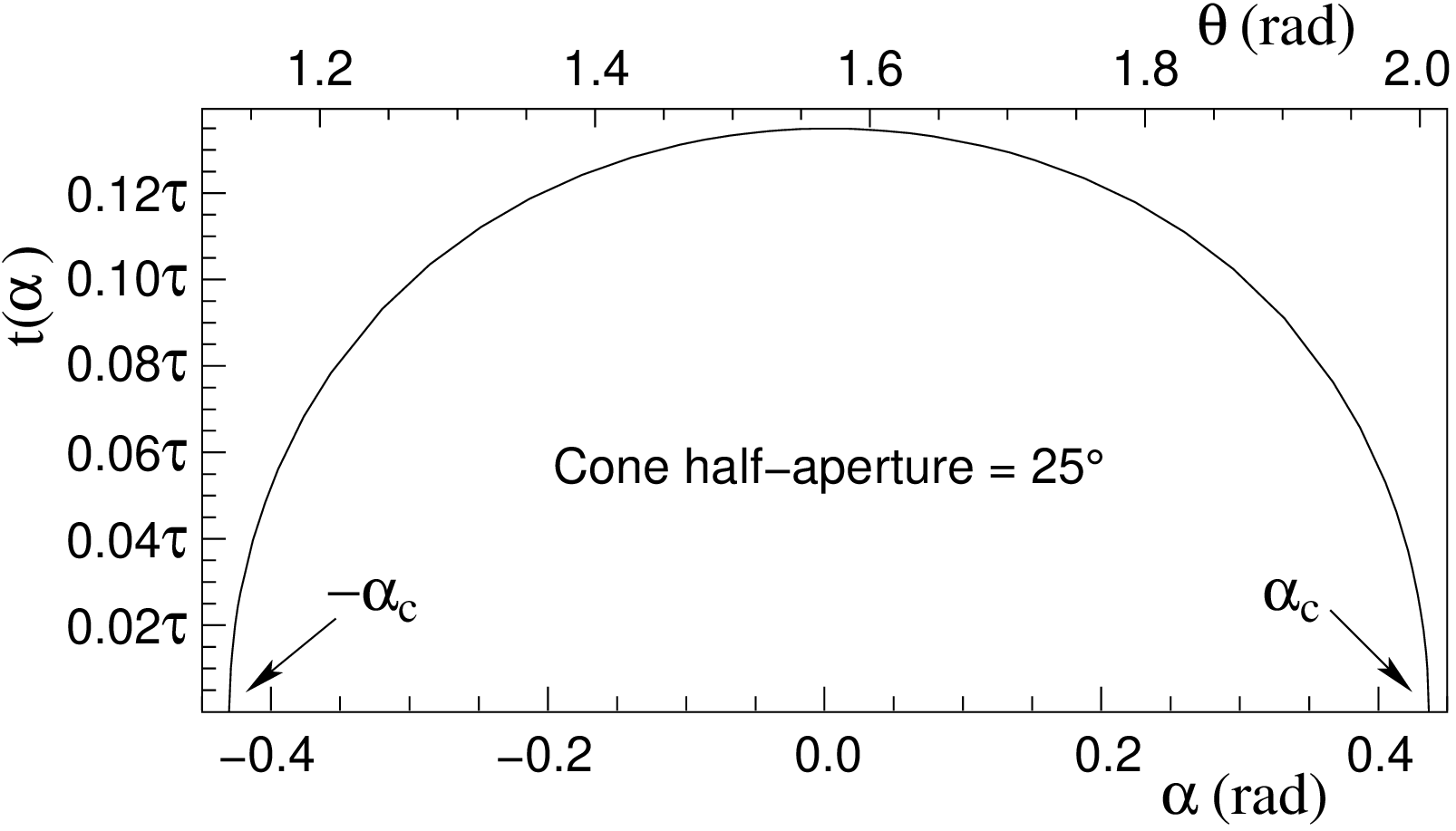}
\caption{Fractional exposure time (in units of the orbit period
  $\tau$) of a cone with $\alpha_\mathrm{c} = 25$ deg, as function
  of the angles $\alpha$ and $\theta$.}\label{figexp}
\end{figure}

 Finally, the time $t(\alpha)$ during which the point P, seen at the
 angle $\alpha$ with respect to the equatorial plane, is viewed by the
 detector is:
\begin{equation}
  t(\alpha) = \frac{\tau}{\pi} \,
    \arcsin \! \sqrt{1 - \frac{\cos^2\alpha_\mathrm{c}}{\cos^2\alpha}}
  \; ,
\end{equation}
 with $-\alpha_\mathrm{c} \le \alpha \le \alpha_\mathrm{c}$
 (see figure~\ref{figexp}, where $\alpha_\mathrm{c} = 25$ deg).

 A better choice of the angle is the orbital colatitude $\theta =
 \frac{\pi}{2} - \alpha$ (also shown in figure~\ref{figexp}): for any
 direction \uvect{d} in the sky one has $\uvect{n} \ps \uvect{d} =
 \cos\theta$.  Every sky element can be mapped using $\theta$ and the
 angle $\phi$ of rotation\footnote{$\phi=0$ define the direction of
 the $x$ axis.}  about \uvect{n}: the solid angle element centered on
 \uvect{d} is $\di\Omega = \sin\theta \di\theta \di\phi$ and its
 exposure per orbit is $t(\uvect{d})\di\Omega$, i.e.:
\begin{equation}\label{eq-t}
  t(\uvect{d})\di\Omega = 
        \dfrac{\tau}{\pi} \, 
    \arcsin \! \sqrt{1 - \dfrac{\cos^2\alpha_\mathrm{c}}{\sin^2\theta}}
        \, \sin\theta \di\theta \di\phi 
\end{equation}
 for $(\frac{\pi}{2}-\alpha_\mathrm{c}) \le \theta \le
 (\frac{\pi}{2}+\alpha_\mathrm{c})$ and $t(\uvect{d})\di\Omega = 0$
 otherwise.  In the following, we will omit to recall that the
 exposure is zero when the condition
 $(\frac{\pi}{2}-\alpha_\mathrm{c}) \le \theta \le
 (\frac{\pi}{2}+\alpha_\mathrm{c})$ is not satisfied.

 In numeric computations, given the partition of the sky in elements
 $\Delta\Omega = \Delta(\cos\theta) \, \Delta\phi$, corresponding to
 uniform binning in $\phi$ and $\cos\theta$, one can produce the
 exposure histogram using equation (\ref{eq-t}), which can be
 rewritten as:
\begin{equation}\label{eq-t2}
  t(\uvect{d}) \, \Delta\Omega = \frac{\tau}{\pi} \, 
    \arcsin \! \sqrt{1 - \frac{\cos^2\alpha_\mathrm{c}}{1-\cos^2\theta}}
     \, \Delta\phi \, \Delta(\cos\theta) \; ,
\end{equation}
 where $\cos\theta$ is the central value of the considered bin, whose
 width is $\Delta(\cos\theta)$.

\subsection{Sky coverage in the precession reference system}

 Equation (\ref{eq-t2}) is valid in any $(x, y, z)$ reference system
 where \uvect{n} is the unit vector defining the $z$-axis.  In order
 to compute the effect of the orbit precession, we move from the ORS
 to the $(x', y', z')$ PRS with a rotation about the $x = x'$
 axis\footnote{We are free to choose the direction of both $x$ and
 $x'$ axes, because the result is independent from this choice.}:
\begin{equation}\label{matrix}
 \mtrx{A} = \left( \begin{matrix}
             1 & 0          & 0 \\
             0 & \cos\beta  & \sin\beta \\
             0 & -\sin\beta & \cos\beta  \\
            \end{matrix} \right) \; .
\end{equation}

 In the PRS the orbit precession simply results in a drift along
 $\phi'$ ($\phi'=0$ along the $x'$ direction, increasing following the
 orbital motion) of the sky elements in the $\cos\theta'$ \textsc{vs.}
 $\phi'$ exposure histogram (i.e.\ in a horizontal shift of the column
 contents).  We consider the sky element $\Delta\phi'
 \Delta(\cos\theta')$ centered around the direction which can be
 written in Cartesian coordinates as
\begin{equation}%\label{}
 \begin{cases}
  x' = \sin\theta' \cos\phi' \\
  y' = \sin\theta' \sin\phi' \\
  z' = \cos\theta' \\
 \end{cases} \; .
\end{equation}

 To apply equation (\ref{eq-t2}), which is independent of $\phi$, we
 only need to find $z = \cos\theta$ by applying $\mtrx{A}^{-1}$,
 obtaining:
\begin{equation}\label{eq-zold}
 \cos\theta = z = \sqrt{1-\cos^2\theta'} \sin\phi' \sin\beta 
        + \cos\theta' \cos\beta \; .
\end{equation}

\begin{figure}[t!]
 \centering\noindent
 \includegraphics[scale=0.6,clip]{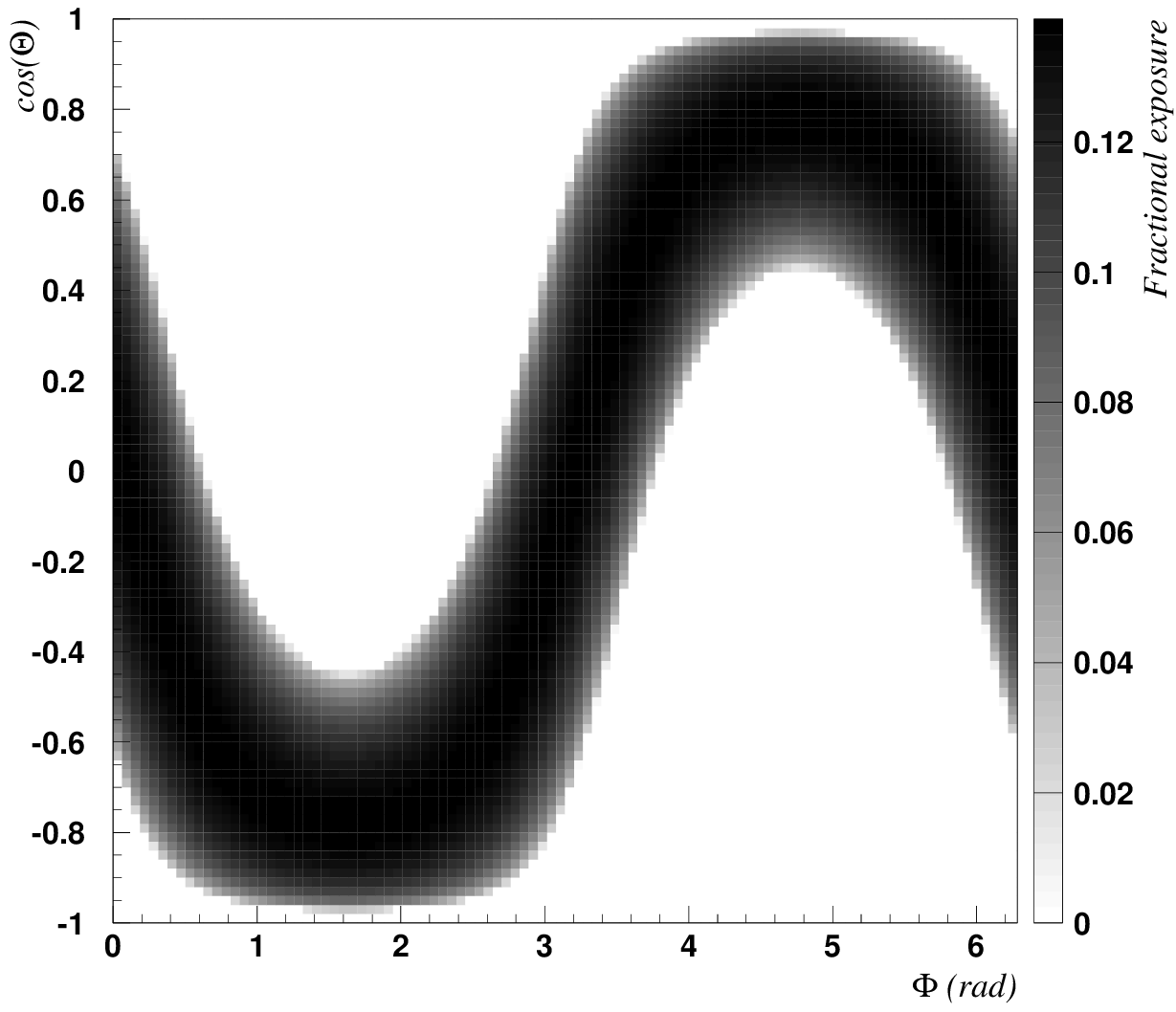}
 \caption[Single orbit exposure map per unit period]{Single orbit
 exposure map per unit period, for a detector with field of view of 50
 deg (25 deg is the cone half aperture), orbiting on board of the the
 ISS.  The \textsf{ExposureFraction} histogram size is $100 \times
 100$ bins.}\label{prs-belt}
\end{figure}

 Using equation (\ref{eq-zold}) together with (\ref{eq-t2}), one can
 obtain a histogram similar to figure~\ref{prs-belt}, which shows the
 sky exposure map (for one orbit, in units of the revolution period)
 seen by a detector whose field of view is a cone with half aperture
 25 deg, installed on the ISS: in the PRS, the ``belt'' is inclined of
 51.6 deg, which is the ISS orbit inclination\footnote{The \uvect{k}
 unit vector points to the North celestial pole, hence one may use the
 celestial coordinates (declination and right ascension) in the PRS.
 This choice fixes the direction of the $x=x'$ axis.}.

 In the following, we will illustrate the method by choosing a mission
 duration $T_\text{mi} = 1000$ days and the detector axis to be
 parallel to the ISS local vertical.  These parameters are similar
 (though not identical) to those of the AMS-02 detector \citep{ams02},
 which will be installed on the ISS in the near future.  The ISS orbit
 parameters are approximated as follows: the revolution period $\tau
 \simeq 1.5$ h, the precession period $T_\text{pr} \simeq 2$ months.
 With altitudes in the range 350--400 km above sea level, the orbit
 eccentricity $e=6.55\times10^{-4}$ is very small (i.e.\ our first
 hypothesis is good).  In addition, the ISS attitude is controlled in
 order to keep its local vertical axis parallel to the radial vector
 (hence the third hypothesis is also good).

\section{Orbit precession}

 Let us call \textsf{ExposureFraction} the histogram representing the
 ``belt'' in the PRS (figure~\ref{prs-belt}), and \textsf{PrsSkyMap}
 the histogram of the exposure in the PRS, which we have now to fill.
 The bin contents of \textsf{ExposureFraction} are pure numbers: they
 must be multiplied by the orbit period $\tau$ to obtain the exposure
 of each $\Delta\Omega$ sky element during a single orbit.

\subsection{Orbit precession in the precession reference system}\label{sec-prs}

 To fill the \textsf{PrsSkyMap} histogram, which has $N_\phi$ bins in
 $\phi'$ and $N_\theta$ bins in $\cos\theta'$, we have to compute the
 effects of the orbit precession, which in the PRS results in a drift
 along $\phi'$ of the sky elements obtained using equation
 (\ref{eq-t2}) after the substitution (\ref{eq-zold}).  The algorithm
 is the following.
\begin{enumerate}
  \item One has first to compute the time required for the orbit
        precession to move each bin on top of the following one.  This
        time is $t_\phi = T_\text{pr} / N_\phi$, with $T_\text{pr} =
        2$ months is the precession period.

  \item Given the mission duration $T_\text{mi}$, one has to find the
        number of times the $\phi$ bins have to be rotated, which is
        $N_\text{ro} = T_\text{mi} / t_\phi$.

  \item Then one has to performs $N_\text{ro}$ iterations which
        consist of a loop over the $N_\phi$ columns, whose single step is
        the sum of the \textsf{ExposureFraction} $(i-1)$-th column
        contents to the \textsf{PrsSkyMap} $i$-th column (with $i$
        decreasing from $N_\phi$ to 1).  Of course, the last column
        should be added to the first one.

  \item Finally, to get the exposure time for each $\Delta\Omega$ sky
        element, one has to multiply each \textsf{PrsSkyMap} bin by
        the revolution period $\tau$.
\end{enumerate}

\begin{figure}[t!]
 \centering\noindent
 \includegraphics[scale=0.6]{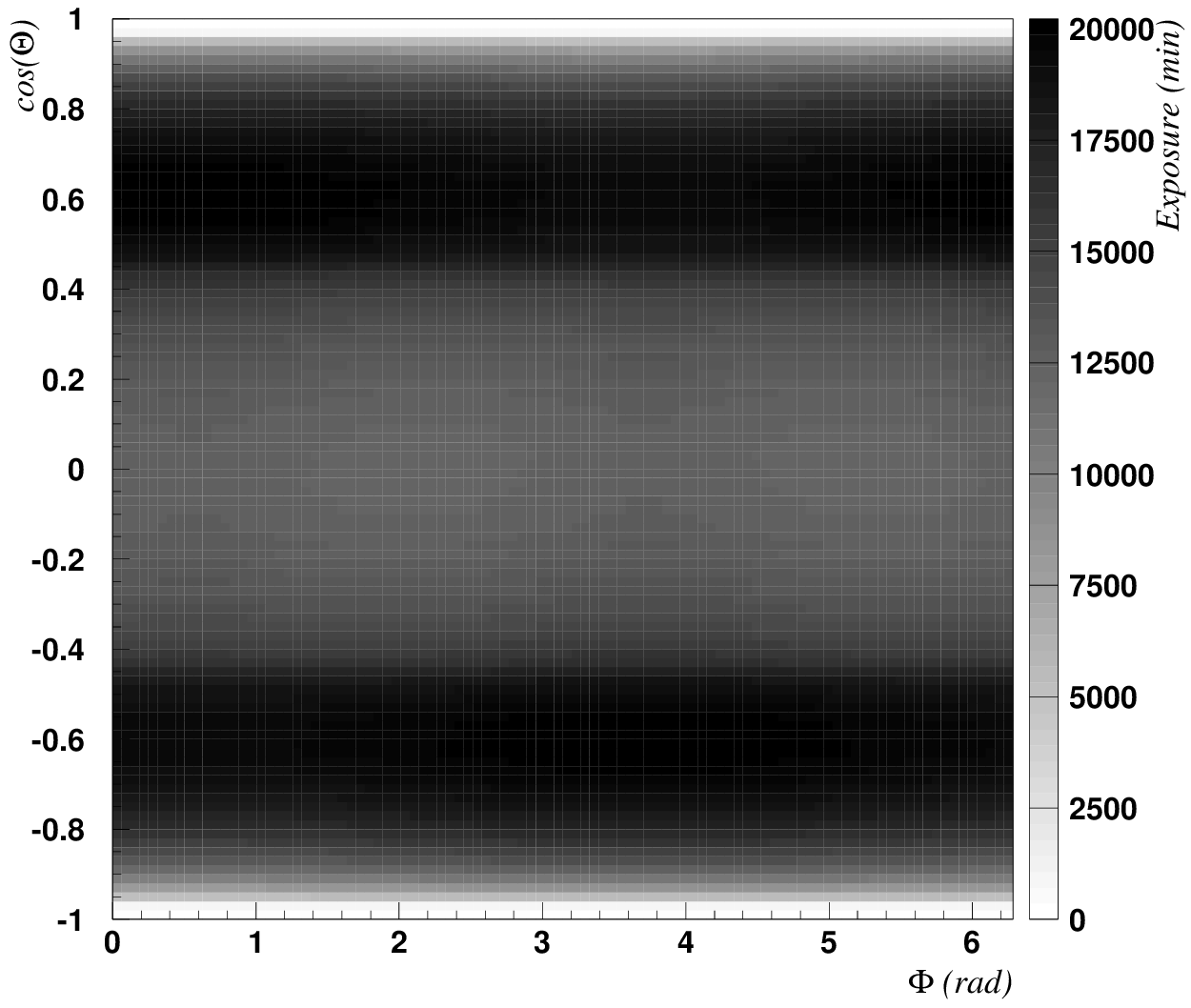}
 \caption[Exposure sky map in the PRS.]{Exposure (in minutes) in the
 PRS of a detector with field of view of 50 deg (25 deg is the cone
 half aperture), orbiting on board of the the ISS.  The orbit period
 is 90 min, the precession period is 2 months, and the mission
 duration is 1000 days.  The \textsf{PrsSkyMap} histogram size is $100
 \times 100$ bins.}\label{prs-prec}
\end{figure}

 Figure~\ref{prs-prec} shows the result for our hypothetical detector
 on the ISS.  The \textsf{PrsSkyMap} histogram has $100 \times 100$
 bins.

\subsection{Orbit precession in the galactic reference system}\label{sec-grs}

 The histogram \textsf{GrsSkyMap} representing the sky map in the GRS
 can be filled with the assumption that the $\Phi$-granularity of
 \textsf{GrsSkyMap} should be identical to the $\phi'$-granularity of
 \textsf{PrsSkyMap} (which is quite reasonable).  However, one can not
 find the PRS histogram and then transform it to the GRS map, because
 continuous transformations can not be applied to discretized maps
 without introducing distortions.  Then the procedure is the
 following.

 For each \textsf{GrsSkyMap} bin, relative to the direction \uvect{d}
 corresponding to the coordinates $(\Phi, \cos\Theta)$ of the bin
 center, we need to find the PRS coordinates which represent the same
 direction:
\begin{equation}
  \uvect{d}_\text{PRS} \equiv 
     (\phi'(\Theta, \Phi), \, \cos\theta'(\Theta, \Phi))
     \equiv (\phi'_j, \, \cos\theta'_i) \; ,
\end{equation}
 where $i \in [0, N_\theta - 1]$ and $j \in [0, N_\phi - 1]$ are the
 row and column indices, respectively, of the \textsf{PrsSkyMap} bin
 which contains $\uvect{d}_\text{PRS}$.

 Each $\phi'$-bin (and $\Phi$-bin, for the previous assumption) is
 $\Delta\phi' = 2\pi/N_\phi$ wide and is inside the detector field of
 view for a time $t_\phi$ per orbit.  In the PRS, the time
 $T(\uvect{d})$ spent inside this bin is the sum over $N_\text{ro}$
 rotations of the \textsf{PrsSkyMap} columns, due to the orbit
 precession.  Hence
\begin{equation}\label{grstime}
  T(\uvect{d}) = \sum_{k=0}^{N_\text{ro}-1}
   \arcsin\sqrt{1-\frac{\cos^2\alpha_\mathrm{c}}{1-F_k(\theta',\phi')}}
\end{equation}
 with
\begin{equation}
 F_{k}(\theta',\phi') =
   \cos^2\theta(\theta'_i,\phi'_j-k\Delta\phi') 
   = \cos^2\theta(\theta'_i,\phi'_{j(k)}) 
\end{equation}
 where $j(k)$ is the $\phi'$-bin which contains
 $\phi'_j-k\Delta\phi'$.  Finally, to obtain the total exposure in the
 considered \textsf{GrsSkyMap} bin, we have to multiply the time spent
 viewing each sky portion centered around $(\Phi, \cos\Theta)$ for
 $N_\text{ro}$ times its solid angle: the result is $N_\text{ro}
 T(\uvect{d}) \Delta\Omega$, where $T(\uvect{d})$ is given by the
 formula (\ref{grstime}).

\begin{figure}[t!]
 \centering\noindent
 \includegraphics[scale=0.6]{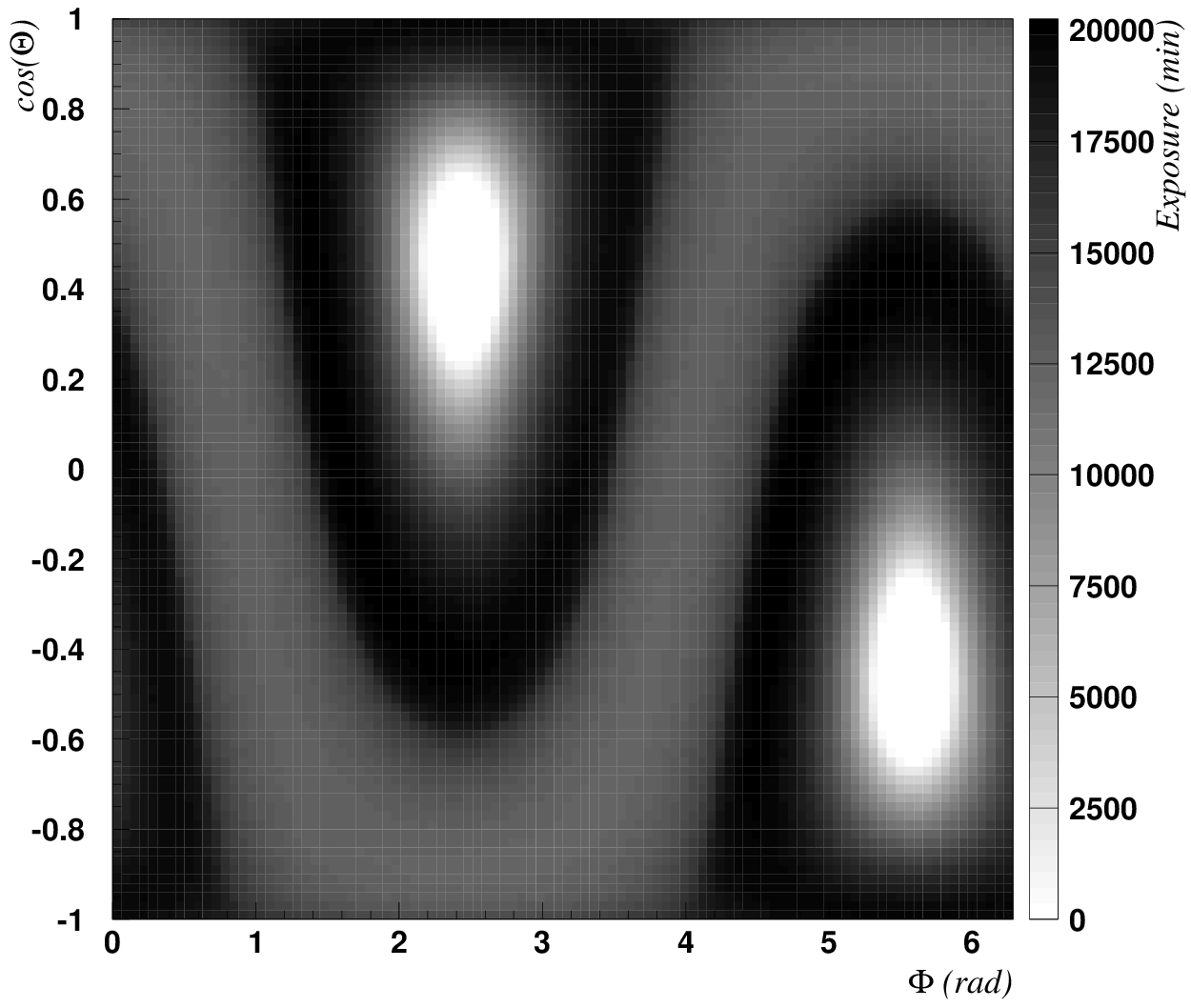}
 \caption{Exposure (in minutes) in the GRS, after an Euler's rotation
 of angles (-33, 62.6, -77.75) deg.}\label{grs-prec}
\end{figure}

 Actually, the real algorithm follow by the program starts from the
 \textsf{GrsSkyMap} histogram and loops over its bins.  For each bin
 in the GRS, it finds the corresponding bin in the PRS using an
 Euler's
 rotation\footnote{\url{http://mathworld.wolfram.com/EulerAngles.html}},
 then applies formula (\ref{grstime}).  Figure~\ref{grs-prec} shows
 the final result, after a rotation with Euler's angles of (-33, 62.6,
 -77.75) deg.  The first angle is the rotation about the third axis,
 which transform the first axis in the `line of nodes'.  The second
 angle is a rotation about the line of nodes, which moves the third
 axis in the final position.  The third angle is a rotation about the
 new third axis, which transforms the line of nodes into the final
 first axis.  In order to show another possible choice of the
 coordinate system, we considered the PRS as the celestial-equatorial
 reference system in our example, and chose the Euler's angles which
 transform it to the GRS, expressed in the B1950.0 epoch as defined in
 1958 by the IAU \citep{zombeck90}.

\section{Detector sensitivity}

 Having obtained the exposure time of each sky element, the detector
 sensitivity is found in the following way.  The first step is to
 compute the product of the effective detector sensitive area $A(E)$
 (usually expressed in cm$^2$) with the sky element exposure
 $\Delta\Omega\, t$ (sr s):
\begin{equation}
  S(E) =  A(E)\, \Delta\Omega\, t \quad (\!\um{cm^2} \um{sr} \um{s}) \; .
\end{equation}
 The effective area $A(E)$ in general will depend on the photon energy
 $E$ and on the incident direction.  Here we consider a uniform
 acceptance over all directions inside the field of view, hence our
 expression for $S(E)$ is equivalent to the average over all
 directions of the real detector efficiency.

 If a known source, covering the sky element $\Delta\Omega$ (which can
 be contained in a single histogram bin or extend over a set of
 adjacent bins), has a photon flux $F(E)$ $(\!\um{cm^{-2}}
 \um{sr^{-1}} \um{s^{-1}})$ in a particular energy bin centered at
 $E$, then the number of events measured by the detector are expected
 to be:
\begin{equation}\label{eq-n-hits}
  N_\text{meas}(E) =  F(E) \, S(E) \; .
\end{equation}

 On the other hand, there will be some background flux of photons
 coming from the same sky element, which will produce background
 events in each energy bin.  Formula (\ref{eq-n-hits}) can also be
 used to find the number of expected background hits in the detector.
 In case one considers a single energy bin, she/he would define the
 detector source sensitivity in this bin as the ratio between the
 number of expected measured events and the square root of the number
 of background hits\footnote{The number of background hits per energy
 bin is assumed to be Poisson distributed.}, or follow the more
 detailed approach of ref.\ \citep{li83}.  However, the usual approach is to
 consider the whole detector energy range at the same time, and to
 perform a fit of the measured distribution (deconvolved with the
 detector sensitivity) with a function which is modeling both the
 source spectrum and the background flux at the same time.

 In this way, for each set of detector parameters, it is possible to
 identify the sources which are visible with the considered orbit and
 mission duration.

\section{Conclusion}

 The analytical method developed in this paper allows for a fast
 estimation of the detector sensitivity to known sources, when the
 hypotheses considered in section~\ref{sec-hyp-alg} are valid.  The
 source code, freely available on the author's web
 site\footnote{\url{http://cern.ch/casadei/software.html}}, is a
 useful tool during the design phase of an experiment making use of an
 orbital detector, when the goal is the definition of its parameters
 and of the mission duration.  In addition, it is useful also as a way
 to easily check the performance of existing detectors, when access to
 their detailed simulation program is not easy or allowed.

 On the other hand, if one really needs a precise estimation of the
 detector performance, she/he would need to make use (or to develop) a
 program which exactly tracks the satellite along its orbital motion
 using complicated formule, which can be found in papers like ref.\
 \citep{murad95} or textbooks like ref.\ \citep{wertz78} and
 \citep{roy88}.  In addition, the program should be able to correctly
 simulate the detector response to photons with different energies and
 incident directions, as was done for example by the AMS collaboration
 \citep{natale01,sevilla04,ams2dmgamma,valle04}.  But the price for
 the higher precision is a CPU intensive and complex program (with
 usually harder user interface than this analytic method), and a long
 development time.

\section*{Acknowledgments}

 The author wishes to thank S. Natale, I. Sevilla-Noarbe and S. Bergia
 for the useful discussions, together with S.S. Shore and V. Vitale
 for their comments and corrections to the first draft of this paper.

\bibliography{coverage}

\end{document}